\documentclass[prl,twocolumn,final,amsmath,amssymb,superscriptaddress,floatfix,showpacs,a4paper,aps,10pt,hidelinks]{revtex4-1}

\usepackage{graphicx,color} 
\usepackage[colorlinks=false]{hyperref}
\usepackage{multibib}
\usepackage{balance}
\usepackage[utf8]{inputenc}

\newcommand{\sz}{$^1\mathrm{S}_0$ }
\newcommand{\pz}{$^3\mathrm{P}_0$ }
\newcommand{\ponez}{$^3\mathrm{P}_1$}

\newcommand{\ket}[1]{\ensuremath{\left|#1\right\rangle}}
\newcommand{\bra}[1]{\langle\left.{#1}\right|}

\begin{document}

\title{Observation of an Orbital Interaction--Induced Feshbach Resonance in $^{173}$Yb}


\author{M. H\"ofer}
\affiliation{Ludwig-Maximilians-Universit\"at, Schellingstra\ss{}e 4, 80799 M\"unchen, Germany}
\affiliation{Max-Planck-Institut f\"ur Quantenoptik, Hans-Kopfermann-Stra\ss{}e 1, 85748 Garching, Germany}
\author{L. Riegger}
\affiliation{Ludwig-Maximilians-Universit\"at, Schellingstra\ss{}e 4, 80799 M\"unchen, Germany}
\affiliation{Max-Planck-Institut f\"ur Quantenoptik, Hans-Kopfermann-Stra\ss{}e 1, 85748 Garching, Germany}
\author{F. Scazza}
\affiliation{Ludwig-Maximilians-Universit\"at, Schellingstra\ss{}e 4, 80799 M\"unchen, Germany}
\affiliation{Max-Planck-Institut f\"ur Quantenoptik, Hans-Kopfermann-Stra\ss{}e 1, 85748 Garching, Germany}
\author{C. Hofrichter}
\affiliation{Ludwig-Maximilians-Universit\"at, Schellingstra\ss{}e 4, 80799 M\"unchen, Germany}
\affiliation{Max-Planck-Institut f\"ur Quantenoptik, Hans-Kopfermann-Stra\ss{}e 1, 85748 Garching, Germany}
\author{\mbox{D. R. Fernandes}}
\affiliation{Ludwig-Maximilians-Universit\"at, Schellingstra\ss{}e 4, 80799 M\"unchen, Germany}
\affiliation{Max-Planck-Institut f\"ur Quantenoptik, Hans-Kopfermann-Stra\ss{}e 1, 85748 Garching, Germany}
\author{M. M. Parish}
\affiliation{School of Physics and Astronomy, Monash University, Victoria 3800, Australia}
\author{J. Levinsen}
\affiliation{School of Physics and Astronomy, Monash University, Victoria 3800, Australia}
\author{I. Bloch}
\affiliation{Ludwig-Maximilians-Universit\"at, Schellingstra\ss{}e 4, 80799 M\"unchen, Germany}
\affiliation{Max-Planck-Institut f\"ur Quantenoptik, Hans-Kopfermann-Stra\ss{}e 1, 85748 Garching, Germany}
\author{S. F\"olling}
\affiliation{Ludwig-Maximilians-Universit\"at, Schellingstra\ss{}e 4, 80799 M\"unchen, Germany}
\affiliation{Max-Planck-Institut f\"ur Quantenoptik, Hans-Kopfermann-Stra\ss{}e 1, 85748 Garching, Germany}
\email{simon.foelling@lmu.de}

\date{\today}


%


\begin{abstract}
We report on the experimental observation of a novel interorbital Feshbach resonance in ultracold $^{173}$Yb atoms. This opens up the possibility of tuning the interactions between the \sz and \pz metastable state, both possessing zero total electronic angular momentum. The resonance is observed at experimentally accessible magnetic field strengths and occurs universally for all hyperfine state combinations. We characterize the resonance in the bulk via interorbital cross thermalization as well as in a three-dimensional lattice using high-resolution clock-line spectroscopy. Our measurements are well described by a generalized two-channel model of the orbital-exchange interactions.
\end{abstract}

\maketitle

Magnetic Feshbach resonances have become an indispensable tool in the study of ultracold quantum gases, enabling the tuning of interaction strengths over a wide parameter range \cite{Chin2010}. 
This tunability has given rise to an impressive set of experimental results, including the realization of the BEC--BCS crossover in degenerate Fermi gases \cite{Zwierlein2004, Regal2004, Chin2004, Zwerger2012} as well as the discovery of novel few-body phenomena such as Efimov trimers \cite{Kraemer2006}.

Whereas the majority of alkali atomic species feature Feshbach resonances, this is not the case for alkaline-earth-like atom states without electronic angular momentum.  In this Letter, we report on the observation of a recently predicted novel type of Feshbach resonance \cite{Zhang2015}, enabling the tuning of interorbital interactions based on the Zeeman shift of different nuclear spin states of the atoms.

In a Feshbach resonance, a bound state of the energetically inaccessible molecular potential (closed channel) couples to the scattering continuum of the energetically allowed entrance channel. 
This coupling dramatically affects the atomic scattering properties whenever the bound state energy is close to the open channel energy. In magnetic Feshbach resonances, an external magnetic field can tune these states into and out of resonance, as they have different magnetic moments such as in the prototypical example of singlet and triplet scattering channels of two alkali atoms \cite{ Timmermans1999, Duine2004, Kohler2006}.

\begin{figure}[t!]
\begin{centering}
\includegraphics[width=1\columnwidth]{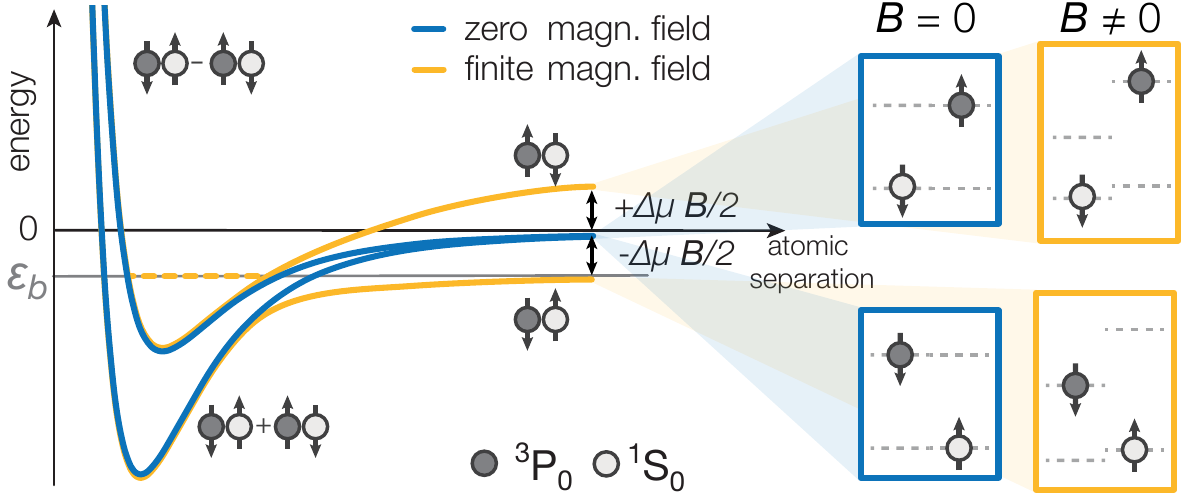}
\par\end{centering}
\protect\caption{Illustration of the magnetic field dependence of the scattering potentials in the two-channel orbital Feshbach resonance. Blue solid lines represent the scattering potentials at zero magnetic field. The electronic triplet (nuclear spin singlet) supports a bound state with energy $\epsilon_b$. A finite magnetic field induces a deformation of the scattering potential for large atomic separation (yellow solid lines), proportional to the differential magnetic moment. The inset illustrates the origin of the differential magnetic moment due to different Landé g-factors of the \sz (bright symbols) and \pz (dark symbols) states.}
\label{fig:schematics}
\end{figure}

In contrast, alkaline-earth-type atoms with two valence electrons such as ytterbium and strontium possess no electronic angular momentum in the atomic ground state, either for the electronic spin singlet \sz (denoted \ket{g}) or for the triplet \pz (\ket{e}). This and the associated suppression of hyperfine coupling make these atoms superb candidates for atomic clocks \cite{Ludlow2015} and for the investigation of new exotic many-body states with extended SU($N$) symmetry \cite{Gorshkov2010}.
However, this also implies that no magnetic Feshbach resonances are expected within the two orbital states.

  Instead, optical Feshbach resonances \cite{Fedichev1996}, based on the coupling to a bound molecular state via a one- or two-photon process have been investigated \cite{Enomoto2008,Yan2013}. 
  Because of their limited tunability and lifetime, such optical Feshbach resonances have been difficult to exploit in experiments.
Until now, magnetic Feshbach resonances in ytterbium have only been observed when atoms are specifically prepared in the $^3$P$_2$ state possessing electronic magnetic moment and hyperfine coupling \cite{Kato2013}.
 However, recent measurements \cite{Scazza2014, Cappellini2014} of the scattering properties of a \sz -- \pz atom pair in two different nuclear spin states suggest the existence of a shallow molecular bound state, leading to the prediction of a magnetically accessible scattering resonance \cite{Zhang2015}.


The scattering resonance is described by interaction channels possessing both orbital and nuclear degrees of freedom in $^{173}$Yb ($I=5/2$).
As in \citet{Scazza2014}, the interaction can be expressed in the basis consisting of symmetric and antisymmetric superpositions $\ket{\pm}=\frac{1}{2}\left(\ket{ge}\pm\ket{eg}\right)\left(\ket{\uparrow\downarrow}\mp\ket{\downarrow\uparrow}\right)$ of the electronic orbitals (\ket{e},\ket{g}) and the two nuclear spin states (\ket{\uparrow},\ket{\downarrow}) with $m_F^\downarrow, m_F^\uparrow\in{-\frac 52,....,+ \frac 52}$.
 Associated to \ket{\pm} are the orbital singlet scattering length $a_\mathrm{eg}^- = 219.5(29)\,a_0$  \cite{Scazza2014} and the very large triplet scattering length $a_\mathrm{eg}^+ >2000$\,$a_0$  \cite{Scazza2014, Cappellini2014}, with $a_0$ denoting the Bohr radius.
 The two orbitals exhibit different Landé g-factors, due to a weak hyperfine coupling
of the \pz with the \ponez ~ state, giving rise to a differential magnetic moment $\Delta \mu = (g_e^{} m_F^\downarrow-g_g^{} m_F^\uparrow)\mu_B$, where $\mu_B$ is the Bohr magneton. The presence of a magnetic field will therefore mix the singlet and triplet states, introducing a coupling between the orbital and spin degrees of freedom.
 In this case, the non-interacting system has the eigenbasis $\ket{o} =(\ket{g\uparrow; e\downarrow}-\ket{e\downarrow; g\uparrow})/\sqrt 2$ and $\ket{c}=(\ket{e\uparrow; g\downarrow}-\ket{g\downarrow; e\uparrow})/\sqrt 2$.
 
 Two atoms entering the potential in the open channel eigenstate \ket{o} couple to \ket{c} through the orbital-mixing interaction term. 
At short distances the molecular interaction potentials dominate and are independent of the magnetic field due to the symmetric nature of the \ket{\pm} states.
Nevertheless, the entrance channel \ket{o} can be brought into resonance with bound states of the interorbital molecular potentials by shifting the entrance energy using the differential Zeeman shift $\Delta\mu B$ as illustrated in Fig.~\ref{fig:schematics}.
The resonance occurs when $\Delta \mu B$ comes close to the binding energy of the bound state in the closed channel $\epsilon_\mathrm{b}\approx-\hbar^2/(m\,a_c^2)$, with $a_c = (a_{eg}^++a_{eg}^-)/2$.


\begin{figure}[t!]
\begin{centering}
\includegraphics[width=1\columnwidth]{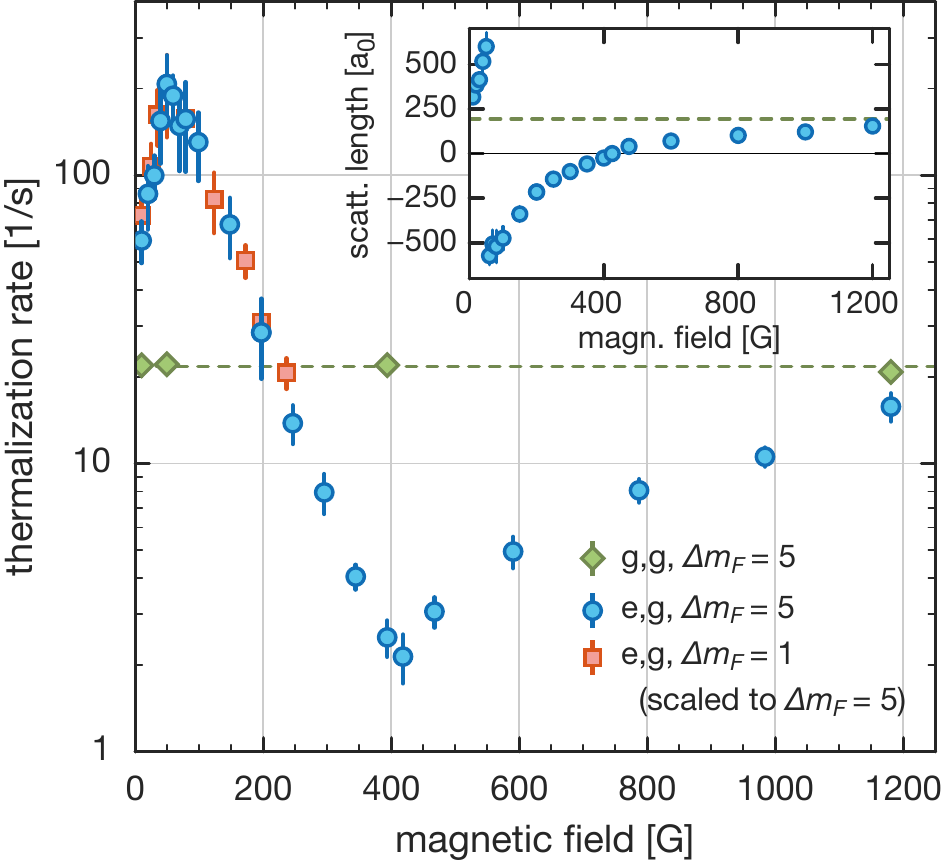}
\end{centering}
\protect\caption{Cross-thermalization rate as a function of the magnetic field $B$ for one- and two-orbital mixtures. Circles and diamonds mark the thermalization rate of the $\ket{e\downarrow} \ket{g\uparrow}$ and the $\ket{g\downarrow} \ket{g\uparrow}$ mixture, respectively with $\Delta m_F = 5$ ($m_F = -5/2,+5/2$). Square symbols show the values for a $\ket{e\downarrow} \ket{g\uparrow}$ mixture with $\Delta m_F = 1$ ($m_F = -5/2, -3/2$) rescaled to $B\rightarrow\frac 15\Delta m_F B$.
Error bars denote the $1\sigma$ uncertainty of the fit to the cloud aspect ratio. 
The inset shows a conversion of the thermalization rate to the scattering length $a_{e\downarrow g\uparrow}$, based on the reference provided by the ground state $a_{gg} = 199.4\,a_0$ \cite{Kitagawa2008} (dashed line). In the inset, the offset due to the residual ground state thermalization rate $\Gamma_{gg,\mathrm{res}}=2.17 s^{-1}$ has been subtracted (see main text).}
\label{fig:CrossTherm}
\end{figure}

To determine the magnetic field dependence of the scattering length, we perform cross-dimensional thermalization measurements in a 3D harmonic trap \cite{Monroe1993}. A mixture of  $\ket{e\downarrow}$ and $\ket{g\uparrow}$  is prepared in an out-of-equilibrium state and rethermalizes with the mean rate of elastic collisions, which can be derived from  Enskog's equation \cite{Goldwin2005}. In the limit of low-energy scattering, the collision rate (and therefore the thermalization rate) is proportional to the elastic scattering cross section $\sigma$.

Our experiments begin by creating a degenerate Fermi gas via evaporative cooling in a crossed dipole trap \cite{Scazza2014}. The desired spin mixture is prepared by optical pumping, resulting in a two-component $\ket{g \downarrow}\ket{g\uparrow}$ gas with typical temperatures $T\simeq 0.2\,T_\mathrm{F}$ and a total atom number of $N_\textrm{a}=3\times10^4$ per spin state, where $T_\mathrm{F}$ is the Fermi temperature.

In order to populate the \pz state, the atoms are first loaded into a one-dimensional, state-independent, i.e., magic-wavelength $\lambda_{\mathrm{m}} =759.3$\,nm lattice \cite{Barber2008} in the Lamb-Dicke regime. This lattice is coaligned with the 579\,nm excitation beam used to apply a $\pi$-pulse on the $\ket{g\downarrow} \longrightarrow \ket{e\downarrow}$ optical clock transition at an applied magnetic field of $B_\textrm{exc}=1200$\,G. Next, the atoms are released adiabatically into a magic-wavelength dipole trap (see supplementary information \cite{SM}).
 The sample is then heated along the $z$-direction by repeated Bragg pulses  using a standing wave along this axis, and the subsequent thermalization of excitation into the orthogonal directions is observed by measuring the cloud aspect ratio $\theta(t)$ after 13\,ms of time of flight. 
We determine the thermalization rate $\Gamma_{eg}$  by a single exponential fit to the time dependence  $\theta(t) = 1+\alpha\exp(-\Gamma_{eg}t)$ \cite{Costa2010}. 

We observe a change in the thermalization rate over 2 orders of magnitude as a function of the magnetic field as shown in Fig.~\ref{fig:CrossTherm}. 
In contrast, for a spin mixture in the ground-state orbital \ket{g}, we find the thermalization rate $\Gamma_{gg}$ to be independent of the magnetic field.
 The resulting $B$-field dependence of $\Gamma_{eg}$ exhibits the characteristic shape of a Feshbach resonance with a peak position $B_0 = 55(8)$\,G, and a zero crossing at $B_\Delta = 417(7)$\,G, both determined by quadratic fits within $\pm 15$\,G ($\pm 40$\,G) regions around the resonance (zero-crossing) position, respectively. 
The excitation process has an efficiency of approximately 90\,\% and the thermalization rate in the zero-crossing regime of $a_{e\downarrow g\uparrow}$ is therefore bounded from below by the thermalization of the $\ket{g \uparrow}$ atoms with the residual $\ket{g \downarrow}$ atoms. 	 

A unique property of this new type of Feshbach resonance is that it involves only two channels, which additionally are fully SU($N$) symmetric in the absence of a magnetic field, making the coupling universal with respect to the choice of  $m_F$ states.
 We demonstrate this by repeating the measurement with another spin mixture with a different value of the differential magnetic moment $\Delta\mu$. When $\Delta \mu$ is taken into account by rescaling the magnetic field axis correspondingly, the data sets collapse onto a single curve without further adjustments (see Fig.~\ref{fig:CrossTherm}), demonstrating the universal behavior with respect to different $m_F$-state combinations.


\begin{figure}
\begin{centering}
\includegraphics[width=1\columnwidth]{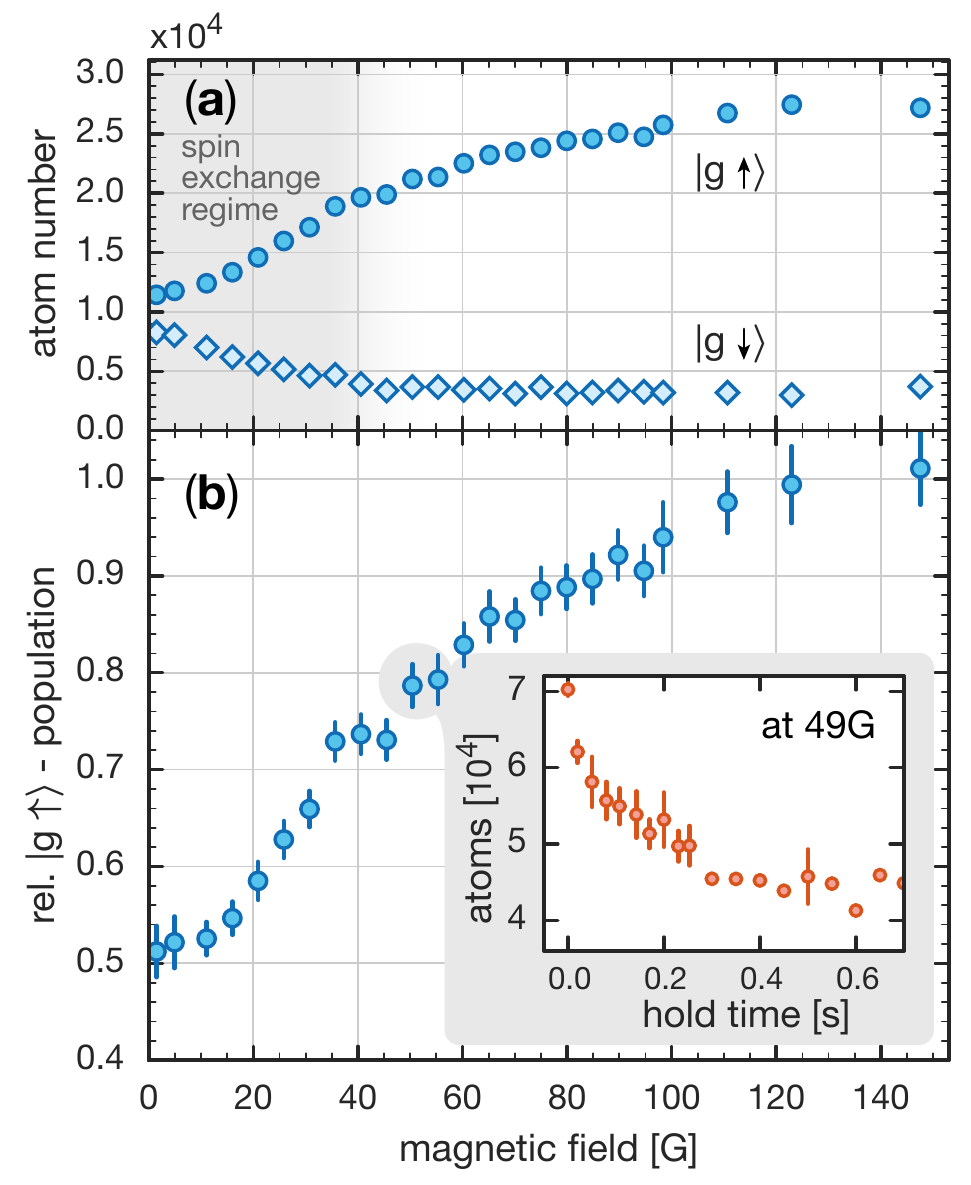}
\end{centering}
\protect\caption{Atom loss in a  $\ket{e\downarrow}\ket{g\uparrow}$ mixture held at a magnetic field $B$ near the orbital Feshbach resonance. (a) Number of atoms in the state \ket{g\uparrow} with $m_F = + 5/2$ (circles) and the residual \ket{g\downarrow} fraction with $m_F = -5/2$ (diamonds). 
 All data points represent averages of at least eight individual measurements and points less than 2.5\, G apart are binned to reduce visual clutter. Error bars indicating the standard error of the mean are smaller than the marker size. 
(b) Relative population of \ket{g\uparrow}, i.e., number of atoms in \ket{g\uparrow} normalized to $\tilde N_{g\uparrow}(B) = \bar N_{g}-N_{g\downarrow}(B)$, where $\bar N_g$ is the ground-state atom number without losses, averaged for fields $B>120$\,G, and $N_{g\downarrow}(B)$ is the residual atom number in \ket{g\downarrow}. 
The inset shows the time trace of the total atom number decay close to the Feshbach resonance.}
\label{fig:Losses}
\end{figure}

In order to characterize the loss channel of the resonance, we use the same preparation protocol as for the cross thermalization, omitting the heating procedure. After a fixed hold time of 150\,ms, the remaining ground-state atoms are imaged spin selectively using an optical Stern-Gerlach technique \cite{Taie2010}.
As shown in Fig.~\ref{fig:Losses}a, when reducing the holding field we observe a magnetic field-dependent loss of the \ket{g\uparrow} population starting close to the position of the resonance. However, the minimum of the loss feature does not occur until the field approaches $B=0$. Below 50\,G, we observe a significant repopulation of the \ket{g\downarrow} state, since orbital exchange becomes energetically favorable at low fields \cite{Scazza2014, Cappellini2014}.
Accounting for this exchange, the loss feature still remains strongly shifted towards lower magnetic fields compared to the resonance position in the elastic channel (Fig.~\ref{fig:Losses}b). This is in fact similar to observations in other mixtures of fermions \cite{Jochim2003, Bourdel2003, Regal2004a,Zhang2011}, but the shift is possibly enhanced due to the intrinsically large size of the shallow bound state. 

At the resonance position, we extract a lifetime of $\tau_{1/e}=386(9)$\,ms at a temperature $T/T_\textrm{F}\simeq 0.3$ and an initial peak density of $n_0 \simeq 5\times 10^{13}$ atoms/$\mathrm{cm}^3$. Assuming three-body decay for the \ket{e\downarrow}\ket{g\uparrow} mixture, we obtain a loss rate coefficient of \mbox{$K_3 = 7.5\times10^{-27}\, {\mathrm{cm^6}}/{\mathrm{s}}$}.  The observed loss is comparable to previously reported values in alkali Fermi gases, e.g., $^{40}$K mixtures \cite{Regal2004a}. 
This realization of a stable strongly interacting Fermi gas appears favorable for exploring novel superfluidity phenomena with this Feshbach resonance in $^{173}$Yb \cite{Zhang2015}.

\begin{figure}[t!]
\begin{centering}
\includegraphics[width=1\columnwidth]{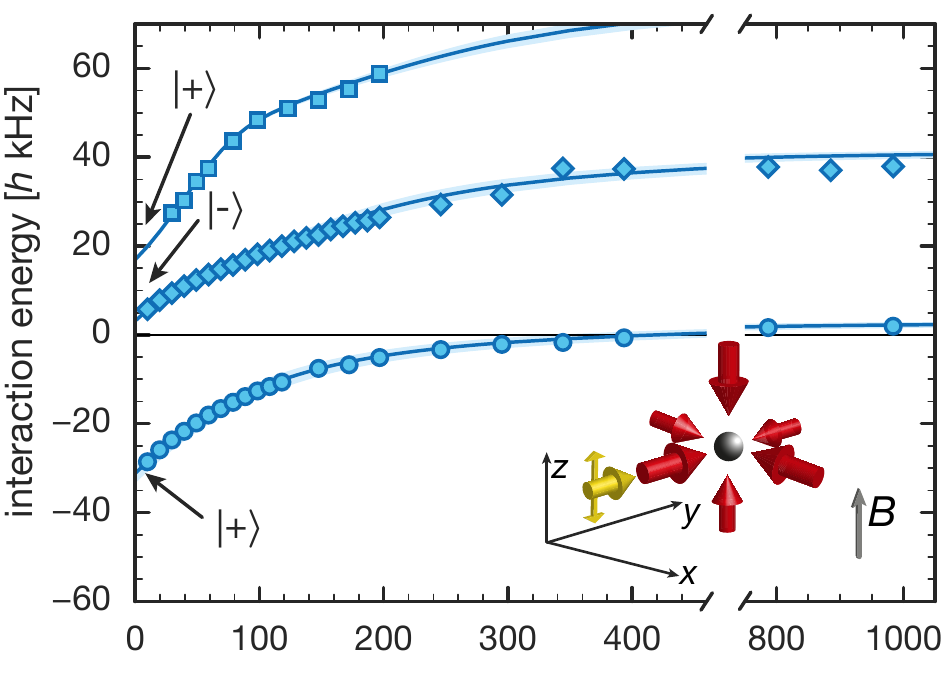}
\par\end{centering}
\protect\caption{Magnetic field dependence of the \ket{e}--\ket{g} atom pair interaction energy on a lattice site. The values are given relative to the energy of two non interacting atoms.
 The assignment of the spectroscopy resonances (points) is based on the observed transition strengths, which decrease for higher lying energy states. 
Solid lines are solutions of the two-particle problem.
Light blue bands indicate the range of variations of the theoretical model spanned by varying $a_{eg}^+$ and $a_{eg}^-$ by 10\,\%. (Inset) Schematic diagram of the excitation geometry.}
\label{fig:OnSite}
\end{figure}

In a third experiment, we directly probe the two-particle interaction on individual sites of a deep, 3D magic-wavelength lattice, as described in \cite{Scazza2014}.
 This offers the advantage of determining the interaction shift of the atom pair with high resolution using clock-line spectroscopy.
  Furthermore, in such a setting the two-particle problem can be separated from possible many-body effects and allows for precise determination of the interaction energies. 
  The energies of the two-particle states that can be excited on the open channel transition at varying magnetic fields are shown in Fig.~\ref{fig:OnSite}. 
  The energy is given relative to that of two non-interacting spatially separated \mbox{atoms, $\epsilon_0$}. 
The data points correspond to resonance positions obtained in an $m_F=\pm5/2$ spin mixture for an isotropic lattice depth of $\tilde{V}=\,29.8 E_\mathrm{r}$, where $E_\mathrm{r}$ is the recoil energy of the lattice light.

The lowest observed branch (circles) has negative interaction energy for $B \lesssim 400$\,G. For low fields, this state corresponds to the molecular branch of \ket{+}.
Extrapolating the measurements to $B=0$, we find an on-site bound state energy of $E_{\textrm{B}}/h=32(2)$\,kHz.

The next higher-lying energy branch was used in \cite{Scazza2014} to determine the value of $a_{eg}^{-}=219.5(29)\,a_0$. We verify with spin-sensitive measurements that this state evolves from the antisymmetric superposition state \ket{-} to the (band-excited) open-channel state \ket{o} with increasing magnetic fields. Using spin-exchange oscillations at low fields, an indirect measurement of $a_{eg}^{+}=3300(300)\,a_0$ was also performed \cite{Cappellini2014}. 

To model the on-site $\ket{e}$-$\ket{g}$ interaction, we generalize the two-channel description of Ref.~\cite{Zhang2015} to the case of isotropic confinement and energy-dependent collisions. In the absence of a magnetic field, the interactions in the singlet and triplet channels decouple. Consequently, the energy shift $E$ is determined by the condition $\sqrt{E+\epsilon_0}\cot\left[\delta_{eg}^\pm(E+\epsilon_0)\right]+\Pi(E)=0$, which depends on the singlet and triplet scattering phase shifts, $\delta_{eg}^\pm$, and the renormalised pair propagator in the lattice site, $\Pi(E)$ \cite{Busch1998,Deuretzbacher2008}. However, in general we need to take into account the coupling introduced by the magnetic field, and this converts the problem into a matrix equation. Thus, the condition for the interaction energy becomes
\begin{equation}
\det[\boldsymbol{\tau}^{-1}(E - \Delta\mu B/2 + \varepsilon_0) +
\boldsymbol{\Pi}(E)] = 0, \label{eq:EnergyCondition}
\end{equation}
with $\boldsymbol{\Pi}(E) = \Pi(E) \ket{o} \bra{\hspace{-2pt}o} + \Pi(E-\Delta\mu B)
\ket{c} \bra{\hspace{-2pt}c} \label{eq:MatrixExpressions}$ and 
$\boldsymbol{\tau}^{-1}(E) = \sqrt{E}\left( \cot\delta^{-}_{eg}(E) \ket{-\hspace{-1pt}}\hspace{-2pt}
\bra{\hspace{-2pt}-} + \delta^{+}_{eg}(E) \ket{+\hspace{-1pt}}\hspace{-2pt} \bra{\hspace{-2pt}+}\right)$.
For details of the derivation, see \cite{SM}.
Note that our model
goes beyond the usual solution of the two-body problem in a harmonic
trap \cite{Busch1998} and accounts for how both open and closed channels are
modified under confinement. We apply a low-energy expansion of the
phase shifts up to and including effective range. The magnetic field
dependence of all measured interaction shifts is very well reproduced
by our model (solid lines in Fig.~\ref{fig:OnSite}). A best fit of the model with $a_{eg}^+$ as a free parameter yields $a_{eg}^+=1878\,\mathrm a_0$. The corresponding effective range $r_\mathrm{eff}^+=216\,  \mathrm a_0$ is
calculated analytically \cite{Gao1998, Flambaum1999}, assuming a long-range van der Waals--potential with $C_6 = 2561$\,{a.u.} \cite{Porsev2014}. The fit uncertainty on $a_{eg}^+$ is $37\,a_0$, but we expect that the uncertainty from model simplifications such as the lattice and scattering phase expansions are comparable or larger than this. 
To indicate the variability, we plot a range corresponding to $\pm10$\% variation of both scattering lengths as shaded areas in Fig.~\ref{fig:OnSite}. The value for $a_{eg}^{-}=219.7\pm 2.2\,a_0$ obtained applying the model to the $B=0$ data from \cite{Scazza2014} is consistent with the one reported there, and has been used in this work with effective range $r_\mathrm{eff}^-=126\,  \mathrm a_0$.

The spectroscopy results also enable us to derive a Feshbach resonance position $B_0 = 50_{7}^{11}$\,G and the zero-crossing $B_\Delta = 327_{62}^{89}$\,G for the bulk at $T=290$\,nK. Both values are in good agreement with the results obtained from the independent cross-thermalization measurements.


In conclusion, we have observed a new type of scattering resonance between different atomic orbitals of $^{173}$Yb arising from strong interorbital spin-exchange interactions. Because of the SU($N$)-symmetric nature of the exchange interaction \cite{Scazza2014}, the resonance occurs universally for any spin combination. We have precisely characterized this Feshbach resonance using a new model of the on-site atom pair interaction energy shift that incorporates the effect of confinement on both open and closed channels. 
Even in the degenerate, strongly interacting quantum gas on resonance, we observe a long lifetime, making our system promising for observing two-orbital Fermi gases with exotic order parameters  \cite{Zhang2015}.
Furthermore, the tunability of the \ket e-\ket g interaction strength suggests novel avenues for the experimental implementation of two-orbital many-body lattice models \cite{Zhang2015a}.

\begin{acknowledgments}
We acknowledge helpful input from Frank Deuretzbacher and valuable discussions with Hui Zhai. This work was supported by the ERC through the synergy grant UQUAM and by the European Union's Horizon 2020 funding (D.R.F.). The work of M.M.P., J.L., and I.B. was performed in part at the Aspen Center for Physics, which is supported by National Science Foundation Grant No. PHY-1066293.
\end{acknowledgments}

%


\renewcommand{\thefigure}{S\arabic{figure}}
 \setcounter{figure}{0}
\renewcommand{\theequation}{S.\arabic{equation}}
 \setcounter{equation}{0}
 \renewcommand{\thesection}{S.\Roman{section}}
\setcounter{section}{0}
\renewcommand{\thetable}{S\arabic{table}}
 \setcounter{table}{0}
 
\onecolumngrid
\newpage

\begin{center}
\noindent\textbf{Supplemental Material}\\
\bigskip
\noindent\textbf{\large{Observation of an Orbital Interaction--Induced Feshbach Resonance in $^{173}$Yb}}

\bigskip
M. H\"ofer$^{1,2}$,  L.Riegger$^{1,2}$, F. Scazza$^{1,2}$, C. Hofrichter$^{1,2}$, D. R. Fernandes$^{1,2}$,\\
M. M. Parish$^{3}$, J. Levinsen$^{3}$, I. Bloch$^{1,2}$ \& S. F\"olling$^{1,2} $
\vspace{0.1cm}

\small{$^1$ \emph{Fakult\"at f\"ur Physik, Ludwig-Maximilians-Universit\"at,\\ Schellingstrasse 4, 80799 M\"unchen, Germany}}

\small{$^2$ \emph{Max-Planck-Institut f\"ur Quantenoptik,\\ Hans-Kopfermann-Strasse 1, 85748 Garching, Germany}}

\small{$^3$ \emph{School of Physics and Astronomy, Monash University, Victoria 3800, Australia}}

\end{center}
\bigskip
\setcounter{page}{1}

\section{Two-channel model in a trap}

To model two $^{173}$Yb atoms of mass $m$ on a single site,
we first consider the two-body problem in a 3D harmonic trap, $V(r) =
\frac{1}2 m\omega_r^2 r^2$. In this case, the center-of-mass and relative
coordinates can be decoupled, and in the relative basis, we have
Hamiltonian $\hat{\mathcal{H}} = \hat{H}_0 + \hat{V}$, with
\begin{align}
\hat{H}_0 = \sum_{n} \epsilon_{n} \ket{o,n}\bra{o, n} +  \sum_{n}
\left(\epsilon_{n} + \Delta\mu B\right) \ket{c,n}\bra{c, n}
\label{Hamiltonian}
\end{align}
and interaction part
\begin{align}
\hat{V} = \sum_{n,n' } \varphi_n(0) \varphi_{n'}(0)  \left\{ U^+ \ket{+,n}
\bra{+,n'} \ + U^- \ket{-,n} \bra{-,n'} \right\}
\end{align}
Here, $n$ labels the relative harmonic oscillator states with angular
momentum $l=0$ (these are the only ones that are affected by the
short-range interactions), $\varphi_n(0)$ is the real-space harmonic
oscillator wavefunction at $r=0$, and
the non-interacting energy $\epsilon_n = 2 n\hbar \omega_r$.
The triplet and singlet configurations are defined as $\ket{\pm} =
\frac{1}{\sqrt{2}} \left( \ket{o} \pm  \ket{c} \right)$, with corresponding
interaction strengths $U^{\pm}$,
which do not depend directly on the nuclear spin.
For simplicity, we focus on a zero-range interaction here, but finite-range
corrections are easily incorporated into the formalism provided the
range of the interactions is much smaller than the harmonic oscillator
length $l_r \equiv \sqrt{\hbar/m\omega_r}$.

To determine the two-body eigenstates, we consider the general wavefunction
in the singlet-triplet basis
\begin{align}
\ket{\psi} = \sum_n \left( b_n^+ \ket{+,n} + b_n^- \ket{-,n}  \right)
\end{align}
and then insert this into the Schr\"odinger equation as follows:
\begin{align}
& \frac{1}{2}
\begin{pmatrix}
1 & 1 \\
1 & -1
\end{pmatrix}
\begin{pmatrix}
\epsilon_n-E & 0 \\
0 & \epsilon_n+\Delta\mu B - E
\end{pmatrix}
\begin{pmatrix}
1 & 1 \\
1 & -1
\end{pmatrix}
\begin{pmatrix}
b_n^+ \\
b_n^-
\end{pmatrix}
+ \varphi_{n}(0)
\begin{pmatrix}
U^+ & 0 \\
0 & U^-
\end{pmatrix}
\sum_{n'}  \varphi_{n'}(0)
\begin{pmatrix}
b_{n'}^+ \\
b_{n'}^-
\end{pmatrix}
= 0
\end{align}
By summing over $n$ and replacing the bare interactions $U^{\pm}$ with the
scattering lengths $a_{eg}^\pm$,
we arrive at the matrix equation for the regular part of the two-channel
wave function $\boldsymbol{\Psi}$:
\begin{align}
\left[\boldsymbol{\tau}_0^{-1} + \boldsymbol{\Pi} (E) \right]
\boldsymbol{\Psi}_{\rm reg} = 0
\end{align}
where
\begin{align}\label{eq:scat}
\boldsymbol{\tau}_0 = - \frac{\sqrt{m}}{\hbar}
\begin{pmatrix}
a_{eg}^+ & 0 \\
0 & a_{eg}^-
\end{pmatrix}
\end{align}
and the quantity
\begin{align} \label{eq:bubble}
\boldsymbol{\Pi}(E) = \frac{\Pi(E)}{2}
\begin{pmatrix}
  1 & 1  \\
1 & 1
\end{pmatrix}
+
\frac{\Pi(E-\Delta\mu B)}{2}
\begin{pmatrix}
1  &  - 1 \\
- 1  &  1
\end{pmatrix}
\end{align}
contains the pair propagator in a harmonic potential \cite{Busch1998Supp}
\begin{align}
\Pi(E) = \frac{\sqrt{2}\hbar}{\sqrt{m}l_r}
\frac{\Gamma(-E/2\hbar\omega_r)}{\Gamma(-E/2\hbar\omega_r-1/2)}
\end{align}
The interaction energies are thus obtained from the condition
$\det\left(\boldsymbol{\tau}_0^{-1} + \boldsymbol{\Pi} (E) \right) =0$.

For finite-range corrections, we need to replace
$\boldsymbol{\tau}_0^{-1}$ by an energy-dependent matrix that depends on the scattering phase shifts, as in Eq.~\eqref{eq:EnergyCondition} of the main text:
\begin{align}
\boldsymbol{\tau}^{-1}(E_c) = \sqrt{E_c}  
\begin{pmatrix}
\cot\delta^{+}_{eg}(E_c) & 0 \\
0 & \cot\delta^{-}_{eg}(E_c)
\end{pmatrix} 
\end{align}
where the collision energy is $E_c=E- \frac{\Delta\mu B}{2} + \frac{3}{2} \hbar\omega_r$.
We are interested in the leading order terms of the low-energy expansion: 
\begin{align}
\frac{\sqrt{mE_c}}\hbar \cot \delta^{\pm}_{eg}(E_c) \simeq - (a^\pm_{eg})^{-1} + \frac{1}{2} r_{\rm eff}^\pm \frac{m E_c}{\hbar^2}
\end{align}
where $r_{\rm eff}^\pm$ denotes the effective range. 
Formally, such an effective range may be included by using a two-channel model
for each of the singlet and triplet interactions.
For the trapped two-body problem, the low-energy expansion of $\boldsymbol{\tau}$ is thus equivalent to making the replacement
$1/a^{\pm}_{eg} \mapsto 1/a^{\pm}_{eg} - \frac{1}{2} m r^\pm_{\rm eff} \left(E- \frac{\Delta\mu B}{2} + \frac{3}{2} \hbar\omega_r\right)$ in Eq.~\eqref{eq:scat}.

The interaction energies of the two-body problem in a 3D harmonic trap are dependent on the trap energy spectrum for the single-channel problem \cite{Busch1998Supp}. In order to take into account the anharmonic character of the lattice potential, we expand this potential to the 6th order and correct the harmonic energy spectrum using second-order perturbation theory. This result matches very well with the diagonalization of the full Hamiltonian performed by Deuretzbacher et. al. \cite{Deuretzbacher2008Supp}.

Using the determined scattering lengths and the corresponding effective ranges of the interactions, one can calculate the scattering amplitude resonance position in the bulk $\tilde B_0 = |\epsilon_\mathrm{b}|/\Delta \mu = 42^{11}_{8}$\,G and the zero crossing $\tilde B_\Delta = 319^{87}_{62}$\,G. These values correspond to zero energy scattering and the error bars were obtained allowing 10\% of uncertainty in the determination of the scattering lengths.

\section{Experimental Methods}
\noindent
In the cross-dimensional thermalization measurement, the atomic cloud is transferred into a crossed dipole trap operating at the magic wavelength with trap frequencies $\omega_{x,y,z} = 2\pi\times(20,120,160)$\,Hz. Simultaneously, the magnetic field is changed in a two-step ramp: First it reaches a value 100\,G above the value of interest in 100\,ms and then is finally ramped with a fixed speed of 2\,G/ms. Directional heating is produced in a controlled way by flashing the lattice potential along the vertical axis. Twenty pulses are applied over a time of three trap oscillation periods, in order to ensure that the cloud is thermalized along the excitation direction. After a variable hold time, the trap is switched off and the atoms in the ground state are imaged after 12\,ms of free flight.

In order to calibrate the magnetic field in our experiments, a measurement of the Zeeman shift of the $\ket{{}^1 \mathrm{S}_0,m_F=\pm 5/2}$ $\rightarrow$ $\ket{{}^3 \mathrm{P}_1,m_F=\pm 7/2}$ transitions between stretched states is performed. A subsequent measurement of the Zeeman shift of the $\ket{{}^1 \mathrm{S}_0,m_F=\pm 5/2}$ $\rightarrow$ $\ket{{}^3 \mathrm{P}_0,m_F=\pm 5/2}$ transitions showed very good agreement with previously published values \cite{Poli2008Supp, Scazza2014Supp}.

\section{Interorbital cross thermalization}
\begin{figure}[h!]
\begin{centering}
\includegraphics[width=1\columnwidth]{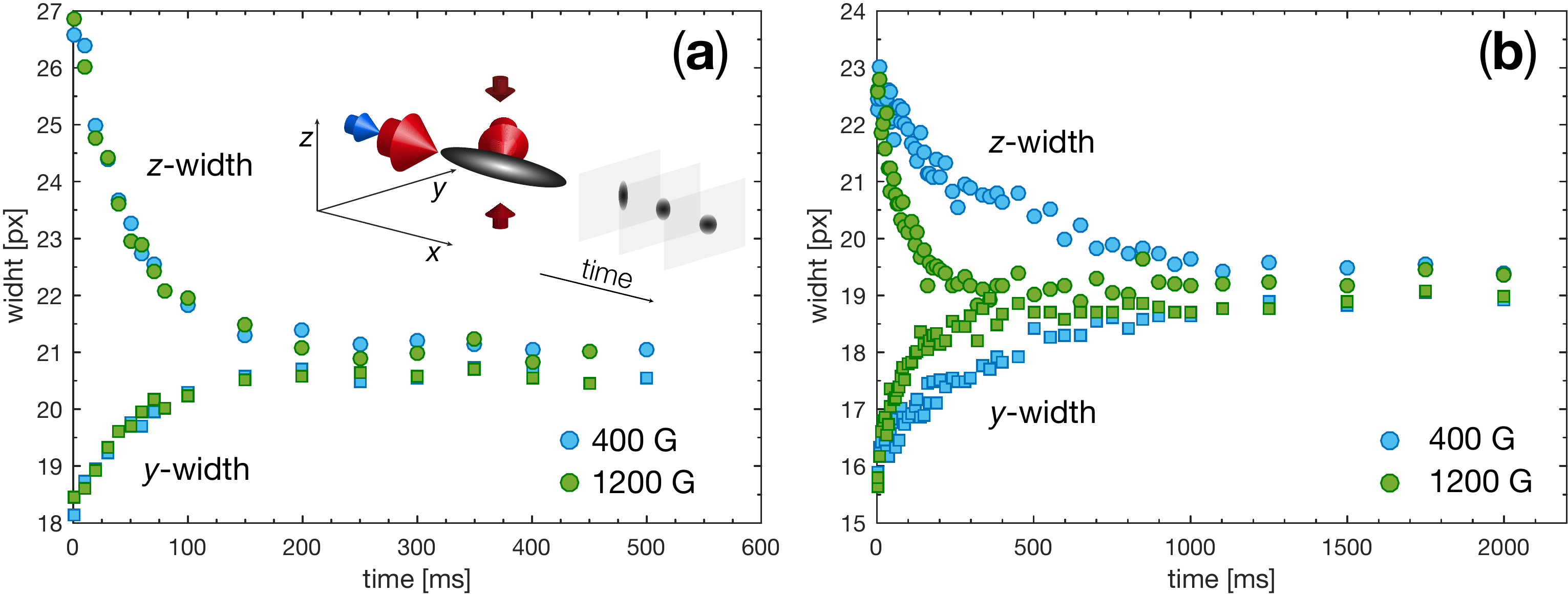}
\end{centering}
\caption{Time evolution of the cloud widths for two different magnetic fields for (a) a mixture of $\ket{g\uparrow} \ket{g\downarrow}$ atoms and (b) a mixture of $\ket{g\uparrow} \ket{e\downarrow}$ atoms. Inset of (a): Trap geometry of thermalization experiment, with dipole trap beams in light red, heating lattice beams in dark red and the imaging beam in blue. The images are oriented in the $yz$-plane.}
\label{figsup:Aspectratio}
\end{figure}

As described in the main text, a cross-dimensional thermalization experiment is performed to determine the magnetic-field dependence of the elastic scattering cross-section of a $\ket{g\uparrow} \ket{e\downarrow}$ mixture.
The aspect ratio of the cloud is measured by performing a Gaussian fit to the density distribution obtained by absorption imaging after 12\,ms of time of flight. As an example, the time evolution of the cloud sizes for two magnetic field values is plotted in Fig.~\ref{figsup:Aspectratio}. The magnetic-field independent thermalization of the $\ket{g\uparrow} \ket{g\downarrow}$ mixture is shown for comparison in \ref{figsup:Aspectratio}(a).

For the cross-thermalization rate measurement, two different values of atomic density are used. For magnetic fields below 200\,G, close to the Feshbach resonance, the density is lower in order to compensate for the large scattering cross-section. At higher magnetic fields, the higher value of atomic density is used. The two data sets are merged together using as reference the corresponding magnetic-field independent thermalization rates of the $\ket{g\uparrow} \ket{g\downarrow}$ mixture.

\end{document}